# Robust OFDM integrated radar and communications waveform design based on information theory

Yongjun Liu*   Guisheng Liao   Zhiwei Yang

National Laboratory of Radar Signal Processing, Xidian University, Xi'an, Shaanxi, 710071, China

*Abstract*- An integrated radar and communications system (IRCS) where a monostatic radar transceiver is employed for target classification while simultaneously used as a communications transmitter is considered. The radar combined propagation-target response (joint response of the radar propagation channel and target) and communications channel response are generally frequency selective but the corresponding frequency response functions are not exactly known. In particular, these frequency response functions are only known to lie in an uncertainty class. To ensure the IRCS simultaneously provides acceptable target classification performance and communications rate, a robust orthogonal frequency division multiplexing (OFDM) integrated radar and communications waveform (IRCW) design method is proposed. The approach finds a waveform that simultaneously provides a sufficiently large weighted sum of the communications data information rate (DIR) and the conditional mutual information (MI) between the observed signal and the radar target over the entire uncertainty class. First, the conditional MI and DIR based on the integrated OFDM radar and communications waveform are derived. Then, a robust OFDM IRCW optimization problem based on the minimax design philosophy is developed such that closed-form solution is derived. Finally, several numerical results are presented to demonstrate the effectiveness of the proposed method.

*Index Terms*- Robust waveform design, integrated radar and communications, orthogonal frequency division multiplexing, conditional mutual information, data information rate

I   INTRODUCTION

With the increased acceptance of communications and radar systems for both commercial and defense applications, the study of integrated systems has attracted significant attention in the signal processing community [1]-[3]. These systems have advantages in reducing the hardware cost and improving the spectrum usage. For example, [4] describes how radar, communications, and electronic

This work was supported by the Foundation for Innovative Research Groups of the National Natural Science Foundation of China [grant No. 61621005]; the National Natural Science Foundation of China [grant No. 61671352]; and China Scholarship Council (CSC).
* Corresponding author.
Y. Liu, G. Liao, and Z. Yang are with the National Laboratory of Radar Signal Processing, Xidian University, Xi'an, Shaanxi, 710071, China (email: yjliuinsist@163.com; liaogs@xidian.edu.cn; yangzw@xidian.edu.cn).

warfare functionality can be integrated into the same platform with array antennas, signal processing, and display hardware shared. The work in [4] uses the advanced multifunction radio frequency concept in order to decrease the system size, weight, and electromagnetic interference while performing multiple functions. However, different functions are carried out by using different waveforms, independently. One important application involves intelligent transportation systems (ITS) [5], [6]. The research in [5] and [6] investigates the integration of radar and communications in an intelligent vehicle system, where the radar can sense collisions and traffic while the communications device connects a vehicle to other vehicles and information sources or collection points. Some different alternative for the design of such systems are surveyed in [3]. In this paper, we will consider an integrated radar and communications system (IRCS) where a monostatic radar transceiver (transmitter and receiver) is employed for target classification while simultaneously used as a communications transmitter.

The integration of radar and communications hardware is a topic of great interest [7]. It seems crucial to explore the best approaches to design the integrated radar and communications waveform (IRCW) that is a single transmitted waveform by the IRCS to perform both radar and communications functions as suggested in [7]. The promising approaches that have already been suggested can be classified into two major categories: multiplexing-waveform and identical-waveform. The multiplexing-waveform approaches employ multiplexing techniques, such as space division multiplexing (SDM) [1], [2], time division multiplexing (TDM) [5], [6], frequency division multiplexing (FDM) [8], and code division multiplexing (CDM) [9]. Multiplexing allows one to easily separate the communications and radar signals so they will not interfere with each other. The identical-waveform approach picks a single waveform which may be similar to the traditional radar waveforms [10] or the conventional communications waveforms [11]. A popular approach uses an orthogonal frequency division multiplexing (OFDM) waveform [12]-[15], a waveform that has been widely applied in communications [16]-[18] and recently suggested for radar applications [19]-[26]. In fact, OFDM waveforms have been proposed for car-to-car (C2C) and car-to-infrastructure (C2I) communications [27]-[29] which is an important application for IRCSs. In this paper, we consider a single pulsed OFDM waveform, called as OFDM IRCW, is transmitted by the IRCS. Such a waveform can simultaneously perform radar and communications functions.

To effectively allocate the limited total power in an IRCS, many design criteria have been proposed. The research in [30] splits the total power between the data symbols that perform the information transmission and the training symbols which accomplish the radar function. The study in [31] and [32] splits the total bandwidth into two subbands, one for communications only and the other for both radar and communications. By employing different allocations of the total power between these two subbands, the joint radar and communications system performance with respect to the data information rate (DIR) and estimation rate is explored. To simultaneously improve the radar target parameter estimation performance and the communications channel capacity, optimal assignment of the subcarrier power profile for an integrated OFDM waveform is proposed in [33] by using a multiobjective design criteria. Similarly, to improve the detection performance and channel capacity of the IRCS, an adaptive transmit power allocation for the subcarriers of an integrated OFDM waveform is proposed in [34]. In fact, there is not a uniform criterion to guide IRCW design.

In this paper, our focus is on a radar used for target classification based on the target-impulse response so our ability to estimate the target-impulse response and thus the minimum mean-square error of that estimate become the appropriate criterion [35]. However, based on [36], this can be shown to be directly related to the maximization of the conditional mutual information (MI) between the observed signal and the radar target return under some reasonable assumptions. This information theoretic criteria [31] [32] has since emerged as a popular criterion for radar waveform design [25], [35]-[37]. On the other hand, information theory provides much of the foundation of communications theory [38], and various communications waveform design methods are proposed to maximize the DIR by properly assigning the total power according to the channel state information (CSI) [39], [40]. Based on this previous work, the research in [41] investigates the conditional mutual information (MI) between the target impulse response and target reflected returns for radar performance while considering channel capacity for communications performance, but waveform optimization is not considered. Based on information theory, the study in [42] explores the adaptive power allocation over the subcarriers of an OFDM IRCW. However, this method needs to know the precise frequency responses of the combined propagation-target and communications channel, which is difficult to attain in practice. In addition, errors in the assumed frequency responses can considerably deteriorate the performance of the IRCS. Hence, it is imperative to explore robust IRCW design method for cases where the frequency responses of the target and communications channel may not be known exactly.

In fact, robust waveform design approaches have been proposed in both radar [43]-[45] and communications [46]-[48] applications. However, robust IRCW design is still an open problem. In this paper, we employ the minimax robust design criterion from [47] to devise a robust IRCW. We assume that the combined propagation-target and communications channel frequency responses are not exactly known, but they lie some uncertainty classes with known upper and lower bounds. A robust OFDM IRCW is developed according to the principle of minimax design that optimizes the worst-case performance of the objective function. The approach devises a waveform that ensures acceptable communications DIR and conditional MI between the observed signal and the radar target return regardless of the actual combined propagation-target and channel impulse responses.

The rest of this paper is organized as follows. In Section II, the radar and communications metrics are formulated. In Section III, the minimax robust waveform design method is proposed. In Section IV, several numerical simulations are presented. Finally, conclusions are drawn in Section V.

## II  PROBLEM DESCRIPTION AND MODELING

In this section, we first present the integrated OFDM signal model of the IRCS. Then the conditional MI and DIR of the IRCW are derived.

### A.  Integrated Radar and Communications Signal Model

In Fig. 1(a), the IRCW employed in this paper is shown, which is a pulse OFDM waveform transmitted by the IRCS to simultaneously perform the radar and communications functions. It is a varied version of the conventional OFDM radar waveform and communications waveform depicted in Fig. 1(b) and Fig. 1(c), respectively. In general, each pulse of the traditional OFDM radar waveform consists of one OFDM symbol without carrying communications information and a cyclic prefix (CP) or guard interval (GI), while the conventional OFDM communications waveform is continuous and includes several OFDM symbols with GI and communications information.

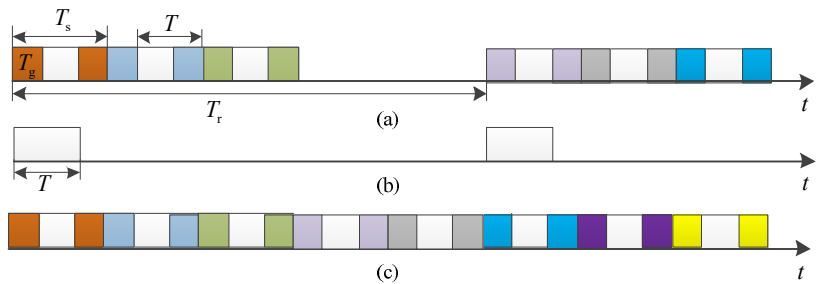

Fig. 1  The transmitted OFDM waveform in (a) OFDM IRCS, (b) OFDM radar, and (c) OFDM communications system.

Following the IRCW shown in Fig. 1(a), the transmitted pulse including $N_s$ consecutive OFDM symbols can be formulated as

$$s(t) = e^{j2\pi f_c t} \sum_{n=0}^{N_s-1} \sum_{m=0}^{N_c-1} a_m c_{m,n} e^{j2\pi m \Delta f (t-nT_s)} \mathrm{rect}\left[(t-nT_s)/T_s\right] \qquad (1)$$

where $f_c$ is carrier frequency, $N_c$ is the number of subcarriers, $\Delta f$ is the subcarrier interval and it satisfies $\Delta f = 1/T$ where $T$ is the elementary length of each OFDM symbol, and the duration of each completed OFDM symbol is $T_s = T + T_g$, where $T_g$ is the duration of GI. For simplicity, the OFDM symbol will refers to the completed OFDM symbol in the following. Moreover, $a_m$ is the transmitted weight over the $m$-th subcarrier, $c_{m,n}$ is the transmitted communications code of the $m$-th subcarrier and $n$-th OFDM symbol to transfer the communications information, and $\mathrm{rect}[t/T_s]$ is the rectangle function, which is equal to one for $0 \leq t \leq T_s$, and zeros, otherwise.

*B. Conditional Mutual Information*

For radar target identification and classification, the target impulse response is normally employed. Hence, it is imperative for the radar to obtain a precise estimation of the target impulse response. This can be guaranteed if there is a sufficiently large conditional MI between the observed signal and the target impulse response [35], called conditional MI in this paper for short. In fact, under some conditions [36] minimizing the mean square error (MSE) in estimating the target impulse response can be equivalent to maximizing the conditional MI. The greater of the conditional MI, the less measurement error of target frequency response will be. In this section, this conditional MI that the IRCS can obtain with the integrated OFDM waveform is derived.

Assume that the impulse response of the propagation channel from transmitter to target and target to receiver is $h_r(t)$, called the radar channel, and that an extended target, as defined in [35], is illuminated, which has an impulse response $g(t)$ which is a Gaussian random process. Therefore, due to the transmitted signal $s(t)$ the received signal of the IRCS is described as

$$y(t) = s(t) * h_r(t) * g(t) + n(t) \qquad (2)$$

where $*$ denotes the convolution operator and $n(t)$ is complex additive white Gaussian noise (AWGN) with zero mean and power spectral density $N(f)$. Note that the delay-Doppler case is not considered here, but the generation for such case is possible, as suggested in [35], while it will complicate the expression.

Suppose that for any frequency $f$ in $\Delta_m = [f_m, f_{m+1}]$, $S(f) \approx S(f_m)$, $H_r(f) \approx H_r(f_m)$, $G(f) \approx G(f_m)$, $N(f) \approx N(f_m)$, where $S(f)$, $H_r(f)$ and $G(f)$ are the Fourier transforms of $s(t)$, $h_r(t)$ and $g(t)$, respectively, and $f_m = f_c + m\Delta f$ is the $m$-th subcarrier frequency. Following the guideline in [35], the conditional MI between the target impulse response and the received echoes can be formulated as

$$I\left(y(t); g(t) | s(t), h_r(t)\right) = \frac{\Delta f T_p}{2} \sum_{m=0}^{N_c - 1} \log_2 \left(1 + \frac{|S(f_m)|^2 |H_r(f_m)|^2 |G(f_m)|^2}{N(f_m) T_p}\right) \quad (3)$$

where $T_p = N_s T_s$ is the pulse duration.

To evaluate $I\left(y(t); g(t) | s(t), h_r(t)\right)$, the $|S(f_m)|^2$ must be calculated. First, the Fourier transform of $s(t)$ is expressed as

$$S(f) = T_s \sum_{n=0}^{N_s - 1} \sum_{m=0}^{N_c - 1} a_m c_{m,n} e^{-j\pi m \Delta f T_s} s_a \left(\pi (f - f_m) T_s\right) e^{-j2\pi (f - f_c)(nT_s - T_s/2)} \quad (4)$$

where $s_a(t) = \sin t / t$. Hence, $U(f) = |S(f)|^2$ can be described as

$$U(f) = T_s^2 \sum_{n=0}^{N_s - 1} \sum_{m=0}^{N_c - 1} \sum_{n'=0}^{N_s - 1} \sum_{m'=0}^{N_c - 1} a_m a_{m'}^* c_{m,n} c_{m',n'}^* e^{-j2\pi (f - f_c)(n - n')T_s} e^{-j\pi (m - m')\Delta f T_s} s_a \left(\pi (f - f_m) T_s\right) s_a \left(\pi (f - f_{m'}) T_s\right) \quad (5)$$

where $(\cdot)^*$ denotes the complex conjugation.

If the communications symbol $c_{m,n}$ is normalized to have unit magnitude and is random with zero mean for any fixed $m$ and $n$ while any code symbols for distinct $(m, n)$ are statistically independent then

$$\mathrm{E}\left[c_{m,n} c_{m',n'}^*\right] = \begin{cases} 1, & m = m', n = n' \\ 0, & \text{else} \end{cases} \quad (6)$$

where $\mathrm{E}[\cdot]$ indicates the expectation operator.

In practice, phase shift keying modulation is widely used in communications, and (6) can be attained through precoding [38], although the communications symbol $c_{m,n}$ is determined by the conveyed information. Using (6), the following can be obtained

$$\mathrm{E}\left[U(f)\right] = T_s^2 N_s \sum_{m=0}^{N_c-1} |a_m|^2 \left[s_a\left(\pi(f-f_m)T_s\right)\right]^2 \tag{7}$$

If the number of subcarriers $N_c$ is sufficient large, $U(f)$ will approach $\mathrm{E}\left[U(f)\right]$ [49]. In practice, the number of subcarriers can be much larger than one hundred, and the following reasonable approximation can be achieved

$$U(f) \approx T_s^2 N_s \sum_{m=0}^{N_c-1} |a_m|^2 \left[s_a\left(\pi(f-f_m)T_s\right)\right]^2 \tag{8}$$

At the frequency $f = f_m$, (8) can also be rewritten as

$$\begin{aligned}U(f_m) &\approx T_s^2 N_s \sum_{m'=0}^{N_c-1} |a_{m'}|^2 \left[s_a\left(\pi(f_m-f_{m'})T_s\right)\right]^2 \\ &= T_s^2 N_s |a_m|^2 + T_s^2 N_s \sum_{m'=0, m'\neq m}^{N_c-1} |a_{m'}|^2 \left[s_a\left(\pi(f_m-f_{m'})T_s\right)\right]^2\end{aligned} \tag{9}$$

If the length of GI $T_g$ is zero, $s_a\left(\pi(f_m - f_{m'})T_s\right)$ will be zero, for $m' \neq m$. However, in practice, the length of GI $T_g$ must larger than the maximum time delay of communications channel. The typical value of $T_g$ is $\frac{T}{4}$, $\frac{T}{8}$, or $\frac{T}{16}$. Hence, $s_a\left(\pi(f_m - f_{m'})T_s\right)$ is not zero, but its value is comparable with the sidelobes of $s_a(t)$, which is far less than 1. Thus, the following approximation is reasonable

$$T_s^2 N_s \sum_{m'=0, m'\neq m}^{N_c-1} |a_{m'}|^2 \left[s_a\left(\pi(f_m-f_{m'})T_s\right)\right]^2 \approx 0 \tag{10}$$

From (9) and (10) the following expression can be achieved

$$U(f_m) \approx T_s^2 N_s |a_m|^2 \tag{11}$$

Substituting (11) into (3) yields

$$\begin{aligned}I\left(y(t); g(t) | s(t), h_r(t)\right) &= \frac{1}{2}\Delta f T_p \sum_{m=0}^{N_c-1} \log_2\left(1 + T_s^2 N_s |a_m|^2 |G(f_m)|^2 |H_r(f_m)|^2 / N(f_m) T_p\right) \\ &= \frac{1}{2}\Delta f T_p \sum_{m=0}^{N_c-1} \log_2\left(1 + p_m v_m\right)\end{aligned} \tag{12}$$

where $p_m = |a_m|^2$, and $v_m = T_s^2 N_s |G(f_m)|^2 |H_r(f_m)|^2 / N(f_m) T_p$ is the channel to noise ratio (CNR)

for the $m$-th subchannel. For simplicity, the conditional MI is referred to as MI in the following.

C. *Data Information Rate*

In the IRCS, both the radar and communications performance is employed to assess the overall utility. Since the DIR is an accepted metric to evaluate the communications performance, the DIR of the IRCS will be calculated in this subsection.

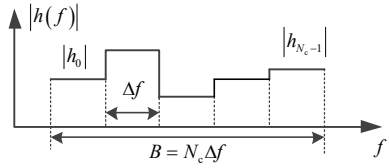

Fig. 2 The frequency response of communications channel.

Without loss of generality, we suppose that the communications channel exhibits slowly varying frequency selective fading. The frequency response of the frequency selective communications channel $h(f)$ is shown in Fig. 2 for a given time. Following the guidelines in [38], the combined DIR from all subchannels can be formulated as

$$C_t = \sum_{m=0}^{N_c-1} \Delta f \log_2\left(1+|a_m|^2 |h_m|^2 / \sigma_c^2\right) = \sum_{m=0}^{N_c-1} \Delta f \log_2\left(1+p_m \varpi_m\right) \tag{13}$$

where $h_m = h(f_m)$ represents the channel frequency response of the $m$-th subchannel, $\sigma_c^2$ is the noise power in the communications channel, $p_m$ effectively indicates the transmit power of the $m$-th subchannel, and $\varpi_m = |h_m|^2 / \sigma_c^2$ can be regarded as the CNR for the $m$-th communications subchannel. From (13), we can see that the higher the CNR is, the larger the DIR.

### III  MINIMAX ROBUST WAVEFORM DESIGN

In this section, we focus on the minimax robust waveform design for the IRCS to optimally allocate the limited transmit power. First, two individual optimal waveform design criteria, i.e., optimal radar waveform design and optimal communications waveform design under known CNR, are discussed.

## A. Radar Waveform and Communications Waveform Design

### 1) Optimal radar waveform for known CNR

In order to improve the performance of radar for target identification and classification, it is imperative to maximize the MI. Hence, with a constraint on total power, the optimization problem can be formulated as

$$\mathbf{p}_r = \arg\max_{\mathbf{p}\in\mathfrak{R}^{N_c}} I\left(y_p(t); g(t)|s(t), h_r(t)\right) = \frac{1}{2}\Delta f T_p \sum_{m=0}^{N_c-1} \log_2\left(1+p_m \upsilon_m\right) \quad (14)$$
$$\text{subject to} \quad \mathbf{1}_{N_c}^T \mathbf{p} \leq 1, \quad p_m \geq 0, m = 0, 1, \cdots, N_c - 1$$

where $\mathbf{1}_{N_c}$ indicates an $N_c \times 1$ vector of ones, $\mathbf{p} = \begin{bmatrix} p_0 & p_1 & \cdots & p_{N_c-1} \end{bmatrix}^T$ is an $N_c \times 1$ vector containing the subcarrier transmit powers, and $\mathbf{p}_r = \begin{bmatrix} p_{r,0} & p_{r,1} & \cdots & p_{r,N_c-1} \end{bmatrix}^T$ is the optimal power allocation for the optimal radar waveform.

The optimization problem is convex, since the object function is concave, and the inequality constraint is convex [50]. The optimal solution can be achieved by utilizing a convex optimization toolbox. In addition, one can obtain a closed-form expression for the solution by using the Karush-Kuhn-Tucker (KKT) conditions [50], and the following optimal solution can be achieved:

$$p_{r,m} = \left[\lambda_r - 1/\upsilon_m\right]^+, m = 0, 1, \cdots, N_c - 1 \quad (15)$$

where $[x]^+ = \max\{x, 0\}$, $\lambda_r$ is the water-level, which satisfies that

$$\sum_{m=0}^{N_c-1}\left[\lambda_r - 1/\upsilon_m\right]^+ = 1 \quad (16)$$

Using the algorithm in [40], the optimal solution in (15) can be specified, and it indicates that more transmit power is assigned to the subchannels with larger CNR, given sufficient power is available.

### 2) Optimal communications waveform for known CNR

For communications, the DIR is an important evaluation criterion and the DIR can be improved through reasonably assignment of the limited transmit power. Under the total transmit power constraint,

the optimization problem can be formulated as follows:

$$\mathbf{p}_c = \arg\max_{\mathbf{p} \in \Re^{N_c}} \sum_{m=0}^{N_c-1} \Delta f \log_2 (1 + p_m \varpi_m) \quad (17)$$
$$\text{subject to} \quad \mathbf{1}_{N_c}^{\mathrm{T}} \mathbf{p} \leq 1, \quad p_m \geq 0, m = 0,1,\cdots,N_c-1$$

where $\mathbf{p}_c = \begin{bmatrix} p_{c,0} & p_{c,1} & \cdots & p_{c,N_c-1} \end{bmatrix}^{\mathrm{T}}$ is an $N_c \times 1$ vector.

The optimization problem in (17) is also convex, since the object function is concave, and the inequality constraint is convex. The following optimal solution can be achieved

$$p_{c,m} = [\lambda_c - 1/\varpi_m]^+, m = 0,1,\cdots, N_c -1 \quad (18)$$

where $\lambda_c$ is the water-level, which satisfies

$$\sum_{m=0}^{N_c-1} [\lambda_c - 1/\varpi_m]^+ = 1 \quad (19)$$

The optimal solution in (18) can be determined by using the algorithm in [40]. It shows that when sufficient power is available, the greater the CNR is in a given subchannel, the more power will be assigned to that subchannel, which is consistent with the optimal solution in (15).

B. *Robust Integrated Radar and Communications Waveform Design*

The individual optimal radar and communications waveforms design have been discussed. For the IRCS, both the radar and communications performance deserve to be considered. In [42], we have proposed the optimal IRCW design method. However, the devised optimal waveform is designed for the specific CNRs which are assumed known. Under such designs the performance may degrade seriously from predicted performance for different CNRs. Furthermore, in practice, it is difficult to know the precise frequency responses of the extended target, radar and communications channels due to the measurement and/or model errors. Hence, it is more reasonable to suppose that the frequency responses of the combined propagation-target and the communications channel are not known exactly, but they lie in the uncertainty classes

$$\Xi_{\text{gh}} = \left\{ \rho_{\text{gh}} : 0 < l_{\text{gh},m} \leq \rho_{\text{gh}}(f_m) = |G(f_m)|^2 |H_r(f_m)|^2 \leq u_{\text{gh},m}, \forall m = 0,1,\cdots, N_c - 1 \right\} \quad (20)$$

$$\Xi_{\text{h}} = \left\{ \rho_{\text{h}} : 0 < l_{\text{h},m} \leq \rho_{\text{h}}(f_m) = |h(f_m)|^2 \leq u_{\text{h},m}, \forall m = 0,1,\cdots, N_c - 1 \right\} \quad (21)$$

for given $l_{\text{gh},m}$, $u_{\text{gh},m}$, $l_{\text{h},m}$, $u_{\text{h},m}$, $\forall m = 0,1,\cdots, N_c - 1$ which might be determined by field measurement or propagation modeling by considering the best and worst cases [45]. We suppose that $l_{\text{gh},m}$, $u_{\text{gh},m}$, $l_{\text{h},m}$, $u_{\text{h},m}$, $\forall m = 0,1,\cdots, N_c - 1$, called the upper and lower bounds, are known.

According to the minimax robust waveform design criterion that optimizes the worst-case performance of the objective function, we wish to solve

$$\max_{\mathbf{p} \in \mathfrak{R}_+^{N_c}} \left\{ \min_{\rho_{\text{gh}} \in \Xi_{\text{gh}}, \rho_{\text{h}} \in \Xi_{\text{h}}} I_{\text{MD}}(\mathbf{p}, \rho_{\text{gh}}, \rho_{\text{h}}) \Big|_{\mathbf{1}_{N_c}^T \mathbf{p} \leq 1} \right\} \quad (22)$$

where $\mathfrak{R}_+^{N_c}$ represents the space $\mathfrak{R}_+^{N_c} = \left\{ \mathbf{x} \big| \mathbf{x} \in \mathfrak{R}^{N_c}, x_m \geq 0, \forall m = 0,1,\cdots, N_c - 1 \right\}$, and

$$I_{\text{MD}}(\mathbf{p}, \rho_{\text{gh}}, \rho_{\text{h}}) = \frac{w_r}{2F_r} \Delta f T_p \sum_{m=0}^{N_c - 1} \log_2\left(1 + p_m \upsilon_m(\rho_{\text{gh}}(f_m))\right) + \frac{w_c}{F_c} \sum_{m=0}^{N_c - 1} \Delta f \log_2\left(1 + p_m \varpi_m(\rho_{\text{h}}(f_m))\right) \quad (23)$$

is the joint performance criterion of the radar and communications system, in which $w_r$ and $w_c$ are weighting factors which satisfy $w_r + w_c = 1$, $\upsilon_m(\rho_{\text{gh}}(f_m)) = N_s T_s^2 \rho_{\text{gh}}(f_m) / (N(f_m) T_p)$, and $\varpi_m(\rho_{\text{h}}(f_m)) = \rho_{\text{h}}(f_m) / \sigma_c^2$. For simplicity, $\upsilon_m$ and $\varpi_m$ represent $\upsilon_m(\rho_{\text{gh}}(f_m))$ and $\varpi_m(\rho_{\text{h}}(f_m))$, respectively. It is an interesting problem to choose the weighting factors, which is related to the relative importance of the radar and communications performance of the IRCS. $F_r$ and $F_c$ are the optimal values (i.e., maximum MI and maximum DIR) in (14) and (17) with the frequency responses of the combined propagation-target and communications channel being the upper bounds. They are normalization factors such that the two performance criteria are approximately within the same range and are of similar magnitudes. The solution to (22) is called the robust waveform design.

To solve (22), we need to obtain the saddle point [46], [47] $\mathbf{p}_S$, $\rho_{\text{gh},S}$, $\rho_{\text{h},S}$ that satisfies

$$I_{\text{MD}}(\mathbf{p}, \rho_{\text{gh},S}, \rho_{\text{h},S})\Big|_{\mathbf{1}_{N_c}^T \mathbf{p} \leq 1} \leq I_{\text{MD}}(\mathbf{p}_S, \rho_{\text{gh},S}, \rho_{\text{h},S})\Big|_{\mathbf{1}_{N_c}^T \mathbf{p}_S \leq 1} \leq I_{\text{MD}}(\mathbf{p}_S, \rho_{\text{gh}}, \rho_{\text{h}})\Big|_{\mathbf{1}_{N_c}^T \mathbf{p}_S \leq 1}. \quad (24)$$

Since $I_{\text{MD}}(\mathbf{p}, \rho_{\text{gh}}, \rho_{\text{h}})$ is monotonically increasing in $\rho_{\text{gh}}(f_m)$ and $\rho_{\text{h}}(f_m)$, the minimum value of

$I_{\text{MD}}(\mathbf{p},\rho_{\text{gh}},\rho_{\text{h}})$ for $\rho_{\text{gh}} \in \Xi_{\text{gh}}$, $\rho_{\text{h}} \in \Xi_{\text{h}}$ is $I_{\text{MD}}(\mathbf{p},\mathbf{l}_{\text{gh}},\mathbf{l}_{\text{h}})$, i.e.,

$$\min_{\rho_{\text{gh}} \in \Xi_{\text{gh}}, \rho_{\text{h}} \in \Xi_{\text{h}}} I_{\text{MD}}(\mathbf{p},\rho_{\text{gh}},\rho_{\text{h}})\Big|_{\mathbf{1}_{N_c}^{\text{T}}\mathbf{p} \leq 1} = I_{\text{MD}}(\mathbf{p},\mathbf{l}_{\text{gh}},\mathbf{l}_{\text{h}})\Big|_{\mathbf{1}_{N_c}^{\text{T}}\mathbf{p} \leq 1} \quad (25)$$

where $\mathbf{l}_{\text{gh}} = \begin{bmatrix} l_{\text{gh},0} & l_{\text{gh},1} & \cdots & l_{\text{gh},N_c-1} \end{bmatrix}^{\text{T}}$, and $\mathbf{l}_{\text{h}} = \begin{bmatrix} l_{\text{h},0} & l_{\text{h},1} & \cdots & l_{\text{h},N_c-1} \end{bmatrix}^{\text{T}}$. Hence, the right most inequality in (24) is satisfied by this choice of $\rho_{\text{gh,S}} = \mathbf{l}_{\text{gh}}$, $\rho_{\text{h,S}} = \mathbf{l}_{\text{h}}$ for any $\mathbf{p} = \mathbf{p}_{\text{S}}$. The left most inequality in (24) requires the solution of $\mathbf{p} = \mathbf{p}_{\text{S}}$ of

$$\max_{\mathbf{p} \in \Re_+^{N_c}} \left\{ I_{\text{MD}}(\mathbf{p},\mathbf{l}_{\text{gh}},\mathbf{l}_{\text{h}})\Big|_{\mathbf{1}_{N_c}^{\text{T}}\mathbf{p} \leq 1} \right\} \quad (26)$$

which is

$$\mathbf{p}_{\text{S}} = \mathbf{p}_{\text{rc}} = \arg\max_{\mathbf{p} \in \Re_+^{N_c}} \frac{w_{\text{r}}}{2F_{\text{r}}} \Delta f T_{\text{p}} \sum_{m=0}^{N_c-1} \log_2(1+p_m v_{l,m}) + \frac{w_{\text{c}}}{F_{\text{c}}} \Delta f \sum_{m=0}^{N_c-1} \log_2(1+p_m \varpi_{l,m}) \quad (27)$$
$$\text{subject to} \quad \mathbf{1}_{N_c}^{\text{T}}\mathbf{p} \leq 1, \quad p_m \geq 0, \, m = 0,1,\cdots,N_c-1$$

where $v_{l,m} = v_m(l_{\text{gh},m}) = N_{\text{s}} T_{\text{s}}^2 l_{\text{gh},m} / (N(f_m)T_{\text{p}})$, and $\varpi_{l,m} = \varpi_m(l_{\text{h},m}) = l_{\text{h},m} / \sigma_{\text{c}}^2$ are determined by the lower bounds. Thus these $\mathbf{p}_{\text{S}}$, $\rho_{\text{gh,S}}$, $\rho_{\text{h,S}}$ satisfy the condition of saddle point conditions.

The objective function in (27) is concave, since it is the affine combination of two concave functions. In addition, the inequality constraint in (27) is convex. Hence, the optimization problem in (27) is convex [48], and it is solvable by using the KKT conditions [48]. Introduce the Lagrange multipliers $\mu$ and $\mu_m$, $m = 0,1,\cdots,N_c-1$, for the constraints in (27). Taking the gradient of the objective function in (27) with respect to $\mathbf{p} = \begin{bmatrix} p_0 & p_1 & \cdots & p_{N_c-1} \end{bmatrix}^{\text{T}}$, setting each component to zero and adding the conditions on $\mu$ and $\mu_m$, for $m = 0,1,\cdots,N_c-1$, from the KKT conditions [48] provides the requirements on the solution to (27) as that $\mathbf{p}$, $\mu$ and $\mu_m$, for $m = 0,1,\cdots,N_c-1$ satisfying

$$\mu - \mu_m = w_{\text{r}} v_{l,m} \Delta f T_{\text{p}} / \left[ 2\ln 2 F_{\text{r}} (1+p_m v_{l,m}) \right] + w_{\text{c}} \varpi_{l,m} \Delta f / \left[ \ln 2 F_{\text{c}} (1+p_m \varpi_{l,m}) \right], \, m = 0,1,\cdots,N_c-1 \quad (28\text{a})$$

$$\mu \left( \sum_{m=0}^{N_c-1} p_m - 1 \right) = 0 \quad (28\text{b})$$

$$\mu_m p_m = 0, \, m = 0,1,\cdots,N_c-1 \quad (28\text{c})$$

$$\mu \geq 0, \mu_m \geq 0, \, m = 0,1,\cdots,N_c-1. \quad (28\text{d})$$

The solution to (28a)-(28d) is

$$p_{\text{rc},m} = \frac{1}{2}\left[\mu'(\alpha'+\beta')-(\upsilon'_{l,m}+\varpi'_{l,m})+\sqrt{\left[(\varpi'_{l,m}-\upsilon'_{l,m})+\mu'(\alpha'-\beta')\right]^2+4\mu'^2\alpha'\beta'}\right]^+ \quad (29)$$

where $\alpha' = w_r \Delta f T_p/(2\ln 2 F_r)$, $\beta' = w_c \Delta f/(\ln 2 F_c)$, $\upsilon'_{l,m} = 1/\upsilon_{l,m}$, $\varpi'_{l,m} = 1/\varpi_{l,m}$, and $\mu' = 1/\mu$ is chosen to solve

$$\left(\sum_{m=0}^{N_c-1} p_{\text{rc},m} - 1\right) = 0 \quad (30)$$

The positive Lagrange multiplier $\mu'$ can be obtained by a simple bisection search over the interval $0 < \mu' \le 1/\min_m\{\alpha'/(\upsilon'_{l,m}+1)+\beta'/(\varpi'_{l,m}+1)\}$, for $m = 0,1,\cdots,N_c-1$, where $\min_m\{x_m\}$ indicates the minimum value in the set $\{x_0,x_1,\cdots,x_{N_c-1}\}$. Once $\mu'$ is obtained, the optimal power assignment will be determined using (29). See Appendix A for the detailed derivations of the optimal solution.

Analyzing the joint performance criterion of the IRCS in (23), we can obtain some useful results that are given by following three theorems.

THEOREM 1: *For a given power allocation, the best radar and communications performance will occur when the true frequency responses are exactly the upper bounds of the uncertainty classes.*

PROOF: From (23), $I_{\text{MD}}(\mathbf{p},\rho_{\text{gh}},\rho_{\text{h}})$ is monotonically increasing in $\upsilon_m(\rho_{\text{gh}}(f_m))$ and $\varpi_m(\rho_{\text{h}}(f_m))$, for $m = 0,1,\cdots,N_c-1$ when the power allocation $\mathbf{p}$ is given. When $\rho_{\text{gh}}(f_m)$ and $\rho_{\text{h}}(f_m)$ are exactly the upper bounds $u_{\text{gh},m}$, and $u_{\text{h},m}$, for $m = 0,1,\cdots,N_c-1$, respectively, both $\upsilon_m(\rho_{\text{gh}}(f_m))$ and $\varpi_m(\rho_{\text{h}}(f_m))$ will be maximum.

THEOREM 2 *For a given power allocation, the worst radar and communications performance will occur when the true frequency responses are exactly the lower bounds of the uncertainty classes.*

Proof is similar to the Proof for Theorem 1.

THEOREM 3: *Suppose that the power allocation is designed to maximize any weighted sum of the DIR*

*and MI assuming some given combined propagation-target frequency response and some communications channel frequency response from the uncertainty classes in (20) and (21). If the combined propagation-target frequency response and the communications channel frequency response both have their minimum values at the same subcarrier and the upper and lower bounds of the uncertainty classes in (20) and (21) are sufficiently separated with respect to the power available, then the worst power allocation puts all the power in this minimizing subcarrier.*

PROOF   See Appendix B for proof.

The peak-to-average power ratio (PAPR) is an important feature of OFDM. The potential high PAPR due to the time varying envelop of OFDM is able to cause the nonlinear distortion of the signal. Due to this, various methods are proposed to reduce the PAPR [51]. Note that in the robust IRCW design only the power assigned over each subcarrier is optimized, hence, in practice, the PAPR reduction techniques without changing the power assignment, such as coding and selected mapping [51], can be used to alleviate the PAPR of the designed robust IRCW.

## IV   SIMULATION

In this section, we present several numerical examples to demonstrate the effectiveness of the proposed robust waveform design method. In the examples, the noise is complex AWGN, and the frequency response of the propagation channel for radar is flat so that $G(\cdot)$ describes the combined propagation-target frequency response. The magnitude of the combined propagation-target frequency response and that of the communications channel frequency response are shown in Fig. 3. Other simulation parameters are shown in Table 1.

Table 1

Simulation parameters

| Parameter | Value | Parameter | Value | Parameter | Value | Parameter | Value |
|---|---|---|---|---|---|---|---|
| GI | 1 us | Number of subcarriers | 128 | Subcarrier spacing | 0.25 MHz | Number of OFDM symbols | 16 |

Since the target impulse response and communications channel response are usually Gaussian functions [35] [52], the scaled Gaussian functions are employed to model the lower and upper bounds of the magnitude of the combined propagation-target and communications channel frequency responses. In Fig. 3, the lower bound of the magnitude of the combined propagation-target frequency response is $\left|G_{\mathrm{L}}(f_m)\right| = e^{-\left[\frac{2(m-N_c/2)}{N_c}\right]^2}$, the upper bound of the magnitude of the combined propagation-target frequency response is $\left|G_{\mathrm{U}}(f_m)\right| = 2 + e^{-\left[\frac{2(m-N_c/2)}{N_c}\right]^2}$, the lower bound of the magnitude of the communications channel frequency response is $\left|h_{\mathrm{L}}(f_m)\right| = e^{-\left[\frac{3(m-N_c/2-30)}{N_c}\right]^2}$, and the upper bound of the magnitude of the communications channel frequency response is $\left|h_{\mathrm{U}}(f_m)\right| = 1.5 + e^{-\left[\frac{3(m-N_c/2-30)}{N_c}\right]^2}$, for $m = 0, 1, \cdots, N_c - 1$.

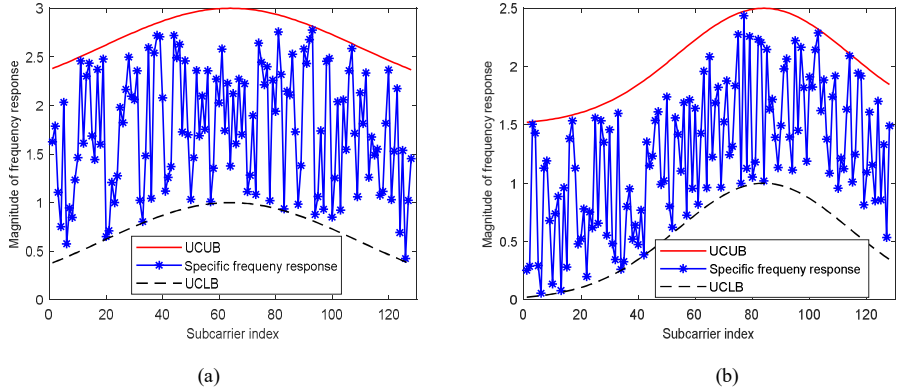

Fig. 3 The uncertainty classes of frequency response. (a) The magnitude of the combined propagation-target frequency response. (b) The magnitude of the communications channel frequency response.

In the following, the non-robust waveform is an optimal waveform designed for the specific frequency responses of combined propagation-target and communications channel in the uncertainty classes as shown in figures. The robust waveform is obtained by solving (22). In the following figures, when the actual combined propagation-target frequency response and communications channel frequency response correspond to the corresponding uncertainty class lower bounds (UCLBs), the performance of the non-robust waveform and that of the robust waveform are labeled as 'NRobW, AFR=UCLB' and 'RobW, AFR=UCLB', respectively. Similarly, when the actual combined propagation-target frequency response and communications channel frequency response correspond to the corresponding uncertainty class upper bounds (UCUBs), the performance of the non-robust waveform and that of the robust waveform are labeled as 'NRobW, AFR=UCUB' and 'RobW, AFR=UCUB', respectively.

A.  Communications Performance

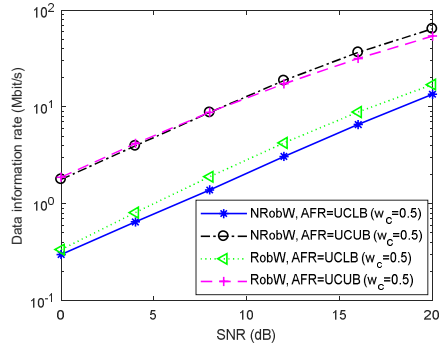

Fig. 4  The data information rate of IRCS.

In Fig. 4, the variation of DIR with the SNR is depicted. The weighting factor for communications is 0.5. The magnitude of the combined propagation-target frequency response and that of the communications channel frequency response are shown in Fig. 3. In Fig. 4, the DIR is enhanced with an increase in the SNR. When the actual combined propagation-target frequency response and communications frequency response correspond to the UCLBs, the robust waveform outperforms the non-robust waveform for the specific frequency responses in Fig. 3, which implies that the worst performance, as described in Theorem 2, of the robust waveform is better than that of the non-robust waveform, since the robust waveform optimizes the worst case performance over the uncertainty class and thus guarantees performance is always better than this quantity. However, the robust waveform is not always superior to the non-robust waveform when the actual combined propagation-target frequency response and communications frequency response correspond to the UCUBs, which indicates that the best performance, as described in Theorem 1, of the robust waveform is not always better than that of the non-robust waveform. As expected, the robust waveform can improve the worst performance of the IRCS.

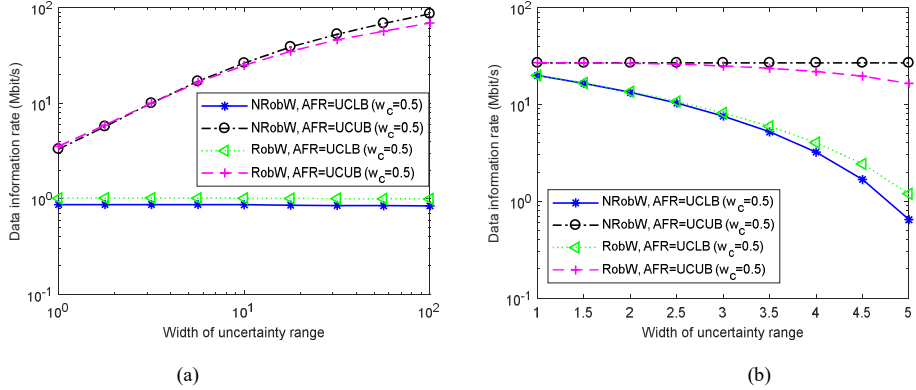

Fig. 5  The variation of data information rate with the width of uncertainty range. (a) The lower bounds are fixed. (b) The upper bounds are fixed.

The dependence of the DIR on the width of the uncertainty range is shown in Fig. 5. In Fig. 5 (a), the lower bounds of the uncertainty classes are fixed, but the upper bounds change with the width of the uncertainty range. The fixed lower bounds of the magnitude of the combined propagation-target frequency response and that of the magnitude of the communications channel frequency response are same as those in Fig. 3. In contrast, in Fig. 5(b), the upper bounds of the uncertainty classes are fixed, but the lower bounds change with the width of the uncertainty range. The fixed upper bounds of the magnitude of the combined propagation-target frequency response is $\left|G_{\mathrm{U}}(f_m)\right| = 5.1 + e^{-\left[\frac{2(m-N_c/2)}{N_c}\right]^2}, m = 0, 1, \cdots, N_c - 1$. The fixed upper bounds of the magnitude of the communications channel frequency response is $\left|h_{\mathrm{U}}(f_m)\right| = 5.1 + e^{-\left[\frac{3(m-N_c/2-30)}{N_c}\right]^2}, m = 0, 1, \cdots, N_c - 1$. The specific frequency responses of communications channel and combined propagation-target are depicted in Fig. 6(a) for fixed lower bounds, and in Fig. 6(b) for fixed upper bounds. The weighting factor for communications is 0.5 and the SNR is 5 dB in Fig. 5.

In Fig. 5(a), when the true frequency responses of combined propagation-target and communications channel correspond to the UCUBs, the performance of the robust waveform and the non-robust waveform is improved as the uncertainty classes become wider with fixed lower bounds. However, when the true frequency responses of combined propagation-target and communications channel correspond to the UCLBs, the performance of the robust waveform and non-robust waveform is unchanged as the uncertainty classes become wider, since lower bounds of the uncertainty classes are fixed. Similar to the results in Fig. 4, the performance of the robust waveform is superior to that of the non-robust waveform when the true frequency responses of combined propagation-target and communications channel correspond to the UCLBs, although the robust waveform is not always better than the non-robust waveform when the true frequency responses of combined propagation-target and communications channel correspond to the UCUBs.

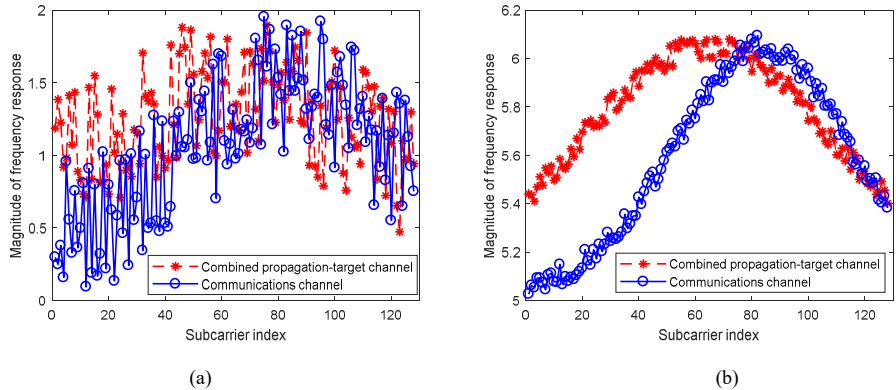

Fig. 6 The magnitude of the specific frequency responses of combined propagation-target and communications channel. (a) The lower bounds are fixed. (b) The upper bounds are fixed.

When the true frequency responses of combined propagation-target and communications channel correspond to the UCLBs, as shown in Fig. 5(b), an increase in the width of uncertainty range causes the performance of the robust waveform and non-robust waveform to be deteriorated, since the lower bounds of the uncertainty classes are decreasing. In contrast, when the true frequency responses of combined propagation-target and communications channel correspond to the UCUBs, the performance of the non-robust waveform is unchanged since the upper bounds of the uncertainty classes are fixed, while the performance of the robust waveform is changed with the decrease of UCLBs, since the robust waveform is determined by the UCLBs under fixed total power. Similar to the results in Fig. 5(a), with the increase of the width of the uncertainty range, the performance of the non-robust waveform eventually becomes worse than that of the robust waveform when the actual frequency responses of the combined propagation-target and communications channel correspond to the UCLBs. As expected, the robust waveform can improve the worst performance of the IRCS.

*B. Radar Performance*

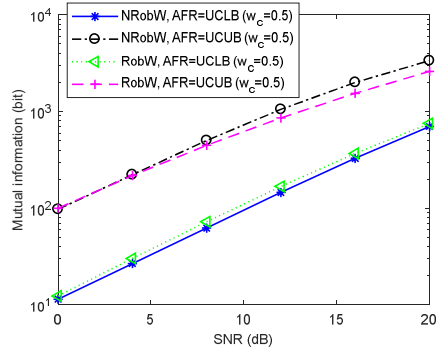

Fig. 7  The mutual information of the IRCS.

In this subsection, the radar performance of the IRCS is evaluated. The dependence of MI on SNR is shown in Fig. 7. The simulation conditions are same as those in Fig. 4. As expected the MI increases with the increase of the SNR. Similar to the communications performance in Fig. 4, the performance of the robust waveform is better than that of the non-robust waveform when the actual frequency responses of combined propagation-target and communications channel equal UCLBs. In this case, both the robust waveform and the non-robust waveform achieve the worst performance as described in Theorem 2. This means that the robust waveform can achieve a favorable performance over the entire uncertainty class, even under the worst possible situation. However, when the true frequency responses of combined propagation-target and communications channel are exactly UCUBs, the performance of the robust waveform is not always better than that of the non-robust waveform.

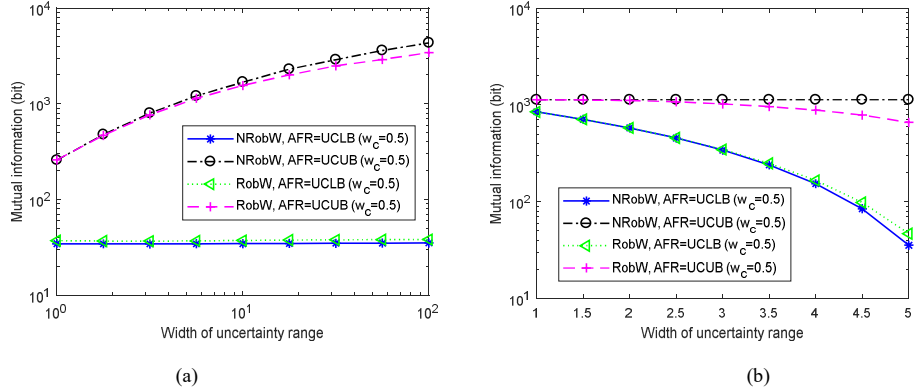

Fig. 8  The variation of mutual information with the width of uncertainty range. (a) The lower bounds are fixed. (b) The upper bounds are fixed.

The variation of the MI with the width of uncertainty range is depicted in Fig. 8. The parameters are same as those in Fig. 5. In Fig. 8(a), the performance of the non-robust waveform and the robust waveform is improved with the increase of the width of uncertainty range when the actual frequency responses of combined propagation-target and communications channel are the UCUBs. At this situation, both the non-robust waveform and robust waveform achieve the best performance as described in Theorem 1. Moreover, the best performance of the robust waveform is not always superior to that of the non-robust waveform. In contrast, when the actual frequency responses of combined propagation-target and communications channel are the UCLBs, the performance of the non-robust waveform and the robust waveform does not change, since the lower bounds of the uncertainty classes stay constant in the example. As expected, the potential worst case performance of the IRCS is improved by using the robust waveform.

In Fig. 8(b), when the true frequency responses of combined propagation-target and communications channel are the UCLBs, the performance of the non-robust waveform and the robust waveform is deteriorated with the increase of the width of uncertainty range, since the lower bounds of the uncertainty classes are decreasing in the example. Moreover, in this case, the performance of the non-robust waveform eventually becomes worse than that of the robust waveform. Similar to the results in Fig. 5(b), when the true frequency responses of combined propagation-target and communications channel are the UCUBs, with the increase of the width of uncertainty range the performance of the non-robust waveform stays constant while that of the robust waveform is gradually deteriorated.

The previous simulation results show that the robust waveform has favorable radar and communications performance. Moreover, the robust waveform can simultaneously improve the worst case performance of the radar and communications, although it is not as good as the non-robust waveform when the true frequency responses of combined propagation-target and communications channel are exactly the UCUBs.

C. Trade-off Curve

For the robust IRCW design, the weighting factors $w_r$ and $w_c$ are required to be specified. In this section we consider how the weighting factors will impact the radar and communications performance of the IRCS by showing all possible solutions for all weighting factors in Fig. 9. In the simulation, the SNR is 15 dB. Fig. 10 shows the lower and upper bounds of the frequency responses of the uncertainty classes along with some specific frequency responses which lie in the uncertainty classes.

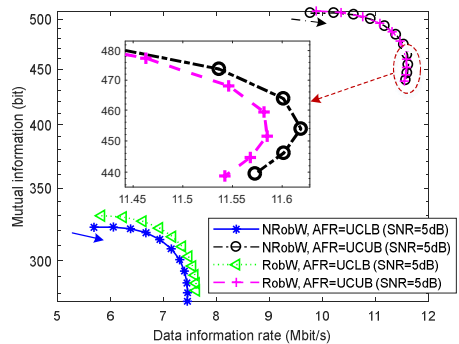

Fig. 9  The optimal trade-off curve of the IRCS.

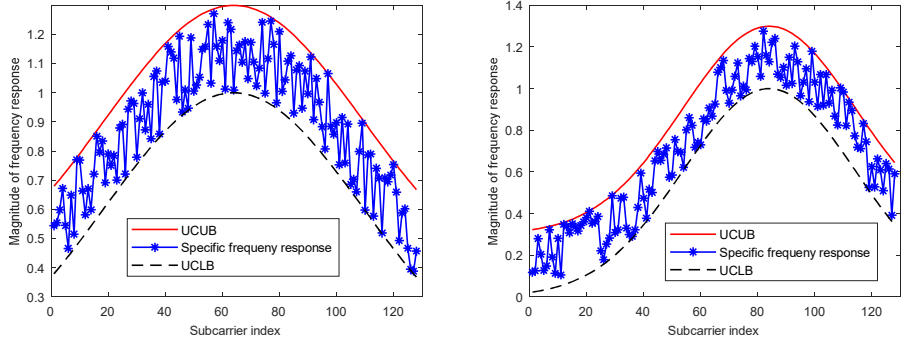

Fig. 10 The uncertainty classes of frequency response. (a) The magnitude of the combined propagation-target frequency response. (b) The magnitude of the communications channel frequency response.

In Fig. 9, in the arrow direction, the weighting factor $w_c$ for communications increases from 0 to 1 in increments of 0.1. As expected, with the increase of the weighting factor for communications, the DIR is improved and the MI is decreased. According to the Theorem 1, the trade-off curves of the robust waveform and the non-robust waveform show the variations of the best performance of these waveforms with the increase of the weighting factor for communications, when the true frequency responses of combined propagation-target and communications channel correspond to the UCUBs. In this case, the trade-off curves show the best trade-off. In contrast, according to the Theorem 2, when the true frequency responses of combined propagation-target and communications channel correspond to the UCLBs, the trade-off curves of the robust waveform and the non-robust waveform show the variations of the worst performance of these waveforms with the increase of the weighting factor for communications. In this case, the trade-off curves show the worst trade-off. For any combined propagation-target and communications channel frequency responses which lie in the uncertainty classes, the performance of the robust waveform will located in the region between the best trade-off curves and the worst trade-off curves. The performance of the non-robust waveform has similar results. Hence, in practice using the trade-off curves one can select the weighting factors to meet the demands for MI and DIR. Moreover, the worst performance of the robust waveform is superior to that of the non-robust waveform although the best performance of the robust waveform is not as good as that of the non-robust waveform due to the fact that the robust waveform can ensure the worst possible performance over the whole uncertainty class is optimal while cannot insure the best possible performance is optimal. The optimal trade-off curves also reveal the inherent compromise between the radar and communications performance, and the trade-off between the robustness and the best possible performance in the IRCW design.

## V Conclusion

In this paper, a robust OFDM IRCW design method is proposed. Under a constraint on the total power, the minimax robust waveform design method is employed to design the robust IRCW and a closed form solution is derived. The devised robust waveform has acceptable performance in the worst case. Moreover, compared with the non-robust waveform the robust waveform has a performance improvement when the true frequency responses of combined propagation-target and communications channel correspond to the UCLBs although it is not as good as the performance of the non-robust waveform when the true frequency responses of combined propagation-target and communications channel correspond to the UCUBs. It is inevitable for the robust OFDM IRCW design to make a trade-off between the radar and communications performance as well as the robustness and the best possible performance. In future work, we will investigate more complicated cases of multiple targets, multiple communications users, color noise, and clutter. The range and Doppler resolution, range sidelobe and communications standard will also be considered in the IRCW design.

## Appendix A
### Derivation of Optimal Solution to the Robust Waveform Design

The KKT conditions in (28) are:

$$\mu - \mu_m = w_r \upsilon_{l,m} \Delta f T_p \Big/ \Big[ 2\ln 2 F_r \big(1 + p_m \upsilon_{l,m}\big) \Big] + w_c \varpi_{l,m} \Delta f \Big/ \Big[ \ln 2 F_c \big(1 + p_m \varpi_{l,m}\big) \Big], m = 0, 1, \cdots, N_c - 1 \text{ (A.1a)}$$

$$\mu\left(\sum_{m=0}^{N_c-1} p_m - 1\right) = 0 \tag{A.1b}$$

$$\mu_m p_m = 0, \ m = 0, 1, \cdots, N_c - 1 \tag{A.1c}$$

$$\mu \geq 0, \ \mu_m \geq 0, \ m = 0, 1, \cdots, N_c - 1 \tag{A.1d}$$

where $\mu$, and $\mu_m$, for $m = 0, 1, \cdots, N_c - 1$, are Lagrange multipliers.

Define $\lambda_m = 1/(\mu - \mu_m)$, $\alpha' = w_r \Delta f T_p / (2 \ln 2 F_r)$, $\beta' = w_c \Delta f / (\ln 2 F_c)$, and (A.1a) can be represented as

$$1/\lambda_m = \alpha' \upsilon_m / (1 + p_m \upsilon_{l,m}) + \beta' \varpi_m / (1 + p_m \varpi_{l,m}) \tag{A.2}$$

Furthermore, (A.2) can be rewritten as

$$1/\lambda_m = \alpha' / (\upsilon'_{l,m} + p_m) + \beta' / (\varpi'_{l,m} + p_m) \tag{A.3}$$

where $\upsilon'_{l,m} = 1/\upsilon_{l,m} > 0$, $\varpi'_{l,m} = 1/\varpi_{l,m} > 0$. Suppose $p_m > 0$, by (A.3) we have

$$p_m^2 + \left[(\upsilon'_{l,m} + \varpi'_{l,m}) - \lambda_m(\alpha' + \beta')\right] p_m + \upsilon'_{l,m} \varpi'_{l,m} - \lambda_m(\alpha' \varpi'_{l,m} + \beta' \upsilon'_{l,m}) = 0 \tag{A.4}$$

The following result can be obtained

$$p_m = \frac{\lambda_m(\alpha' + \beta') - (\upsilon'_{l,m} + \varpi'_{l,m}) \pm \sqrt{\left[(\varpi'_{l,m} - \upsilon'_{l,m}) + \lambda_m(\alpha' - \beta')\right]^2 + 4\lambda_m^2 \alpha' \beta'}}{2} \tag{A.5}$$

According to (A.3), we can obtain that

$$1/\lambda_m < \alpha' / \upsilon'_{l,m} + \beta' / \varpi'_{l,m} \tag{A.6}$$

Equation (A.6) is equivalent to

$$\upsilon'_{l,m} \varpi'_{l,m} - \lambda_m(\alpha' \varpi'_{l,m} + \beta' \upsilon'_{l,m}) < 0 \tag{A.7}$$

Using (A.7) and the property of quadratic equation, we can obtain that the negative root in (A.5) is less than zero, hence

$$p_m = \frac{\lambda_m(\alpha' + \beta') - (\upsilon'_{l,m} + \varpi'_{l,m}) + \sqrt{\left[(\varpi'_{l,m} - \upsilon'_{l,m}) + \lambda_m(\alpha' - \beta')\right]^2 + 4\lambda_m^2 \alpha' \beta'}}{2} \tag{A.8}$$

If $p_m > 0$, in order to satisfy (A.1c), we must have $\mu_m = 0$. Therefore, (A.8) can be rewritten as

$$p_m = \frac{\mu'(\alpha'+\beta') - (\upsilon'_{l,m}+\varpi'_{l,m}) + \sqrt{\left[(\varpi'_{l,m}-\upsilon'_{l,m})+\mu'(\alpha'-\beta')\right]^2 + 4\mu'^2\alpha'\beta'}}{2} \tag{A.9}$$

where $\mu' = 1/\mu$.

If $p_m = 0$, and $\lambda_m = 1/(\mu-\mu_m) \leq 0$, to satisfy (A.3) we must have

$$p_m < 0 \tag{A.10}$$

which is contrary to the practical problem.

If $p_m = 0$, and $\lambda_m = 1/(\mu-\mu_m) > 0$, since $\lambda_m = 1/(\mu-\mu_m) \geq \mu' > 0$, to satisfy (A.3) we have

$$p_m = \frac{\mu'(\alpha'+\beta') - (\upsilon'_{l,m}+\varpi'_{l,m}) + \sqrt{\left[(\varpi'_{l,m}-\upsilon'_{l,m})+\mu'(\alpha'-\beta')\right]^2 + 4\mu'^2\alpha'\beta'}}{2} < 0 \tag{A.11}$$

The results in (A.9) and (A.11) can be summarized as

$$p_{\text{rc},m} = \frac{1}{2}\left[\mu'(\alpha'+\beta') - (\upsilon'_{l,m}+\varpi'_{l,m}) + \sqrt{\left[(\varpi'_{l,m}-\upsilon'_{l,m})+\mu'(\alpha'-\beta')\right]^2 + 4\mu'^2\alpha'\beta'}\right]^+ \tag{A.12}$$

where $[x]^+ = \max\{x,0\}$, and $\mu'$ satisfies that

$$\left(\sum_{m=0}^{N_c-1} p_{\text{rc},m} - 1\right) = 0 \tag{A.13}$$

Hence, the optimal solution to the optimization problem in (27) is obtained.

## APPENDIX B

### PROOF OF THEOREM 3

Without loss of generality, assume that $N_c = 2$, $\mathbf{p}_{\min} = [P_t \ 0]^T$, $\mathbf{p} = [p_1 \ p_2]^T$, $\rho_{\text{gh}}(f_1) \leq \rho_{\text{gh}}(f_2)$ and $\rho_{\text{h}}(f_1) \leq \rho_{\text{h}}(f_2)$, i.e., $\upsilon_1 \leq \upsilon_2$ and $\varpi_1 \leq \varpi_2$, where $p_1 + p_2 = P_t$. We first prove that $I_{\text{MD}}(\mathbf{p}_{\min}, \rho_{\text{gh}}, \rho_{\text{h}}) \leq I_{\text{MD}}(\mathbf{p}, \rho_{\text{gh}}, \rho_{\text{h}})$.

From (23),

$$\begin{aligned}I_{\text{MD}}(\mathbf{p}_{\min}, \rho_{\text{gh}}, \rho_{\text{h}}) &= \alpha\log_2(1+P_t\upsilon_1) + \beta\log_2(1+P_t\varpi_1) \\ &= \alpha\log_2(1+p_1\upsilon_1+p_2\upsilon_1) + \beta\log_2(1+p_1\varpi_1+p_2\varpi_2)\end{aligned} \tag{B.1}$$

where $\alpha = w_r \Delta f T_p/(2F_r)$ and $\beta = w_c \Delta f/F_c$. Now consider a general power weighting

$$\begin{aligned}
I_{\text{MD}}(\mathbf{p},\rho_{\text{gh}},\rho_{\text{h}}) &= \alpha\left(\log_2(1+p_1\upsilon_1)+\log_2(1+p_2\upsilon_2)\right) + \beta\left(\log_2(1+p_1\varpi_1)+\log_2(1+p_2\varpi_2)\right) \\
&= \alpha\log_2(1+p_1\upsilon_1)(1+p_2\upsilon_2) + \beta\log_2(1+p_1\varpi_1)(1+p_2\varpi_2) \\
&= \alpha\log_2(1+p_1\upsilon_1+p_2\upsilon_2+p_1p_2\upsilon_1\upsilon_2) + \beta\log_2(1+p_1\varpi_1+p_2\varpi_2+p_1p_2\varpi_1\varpi_2)
\end{aligned} \quad (\text{B.2})$$

Since $p_1,p_2,\upsilon_1,\upsilon_2 \geq 0$ and $p_1,p_2,\varpi_1,\varpi_2 \geq 0$,

$$I_{\text{MD}}(\mathbf{p}_{\min},\rho_{\text{gh}},\rho_{\text{h}}) \leq I_{\text{MD}}(\mathbf{p},\rho_{\text{gh}},\rho_{\text{h}}) \quad (\text{B.3})$$

For $N_c > 2$, we can obtain the same result. Here we have proved that the worst power allocation puts all the power in the minimizing subcarrier at which both the combined propagation-target frequency response and the communications channel frequency response have their minimum values.

Next, we prove that under some conditions the worst power allocation $\mathbf{p}_{\min}$ also maximizes some weighted sum of the DIR and MI for some given combined propagation-target frequency response and some communications channel frequency response from the uncertainty classes in (20) and (21). These conditions on the frequency responses of combined propagation-target and communications channel will be derived in the following.

Suppose that the worst power allocation $\mathbf{p}_{\min}$ also maximizes some weighted sum of the DIR and MI for some given combined propagation-target frequency response and some communications channel frequency response from the uncertainty classes in (20) and (21), and that the $m_1$-th element of $\mathbf{p}_{\min}$ is 1 and any other element is zero. Hence, $\mathbf{p}_{\min}$ satisfies the optimization solution in (29), i.e.,

$$p_{\text{rc},m_1} = \frac{1}{2}\left[\mu'(\alpha'+\beta')-(\upsilon'_{m_1}+\varpi'_{m_1})+\sqrt{\left[(\varpi'_{m_1}-\upsilon'_{m_1})+\mu'(\alpha'-\beta')\right]^2+4\mu'^2\alpha'\beta'}\right] = P_t = 1 \quad (\text{B.4})$$

$$p_{\text{rc},m_n} = \frac{1}{2}\left[\mu'(\alpha'+\beta')-(\upsilon'_{m_n}+\varpi'_{m_n})+\sqrt{\left[(\varpi'_{m_n}-\upsilon'_{m_n})+\mu'(\alpha'-\beta')\right]^2+4\mu'^2\alpha'\beta'}\right] \leq 0,$$

$$m_n = 0,1,\cdots,N_c-1, \ m_n \neq m_1 \quad (\text{B.5})$$

where $\upsilon'_{m_n} = 1/\upsilon_{m_n}(\rho_{\text{gh}}(f_{m_n})) = (N(f_{m_n})T_p)/N_sT_s^2\rho_{\text{gh}}(f_{m_n})$, and

$\varpi'_{m_n} = 1/\varpi_{m_n}(\rho_{\text{h}}(f_{m_n})) = \sigma_c^2/\rho_{\text{h}}(f_{m_n})$, for $m_n = 0,1,\cdots,N_c-1$.

Simplifying (B.4), the following can be obtained

$$\mu = \frac{1}{\mu'} = \frac{\alpha' + \varpi'_{m_1}\alpha' + \beta' + \upsilon'_{m_1}\beta'}{1 + \upsilon'_{m_1}\varpi'_{m_1} + \upsilon'_{m_1} + \varpi'_{m_1}} \tag{B.6}$$

Simplifying (B.5) we can get

$$\left(\frac{\alpha'}{\upsilon'_{m_n}} + \frac{\beta'}{\varpi'_{m_n}}\right) \leq \frac{1}{\mu'} = \mu, \ m_n = 0,1,\cdots,N_c - 1, \ m_n \neq m_1 \tag{B.7}$$

Substituting (B.6) into (B.7) yields

$$\max_{m_n}\left\{\frac{\alpha'}{\upsilon'_{m_n}} + \frac{\beta'}{\varpi'_{m_n}}\right\} \leq \frac{\alpha' + \varpi'_{m_1}\alpha' + \beta' + \upsilon'_{m_1}\beta'}{1 + \upsilon'_{m_1}\varpi'_{m_1} + \upsilon'_{m_1} + \varpi'_{m_1}}, \ m_n = 0,\cdots,m_1-1, m_1+1,\cdots,N_c-1 \tag{B.8}$$

where $\max_m\{x_m\}$, $m = 0,1,\cdots,N_c-1$, indicates the maximum value in the set $\{x_0, x_1, \cdots, x_{N_c-1}\}$.

Eq. (B.8) can be rewritten as

$$\max_{m_n}\left\{\frac{\alpha'}{\upsilon'_{m_n}} + \frac{\beta'}{\varpi'_{m_n}}\right\} \leq \frac{\alpha'}{1+\upsilon'_{m_1}} + \frac{\beta'}{1+\varpi'_{m_1}}, \ m_n = 0,\cdots,m_1-1, m_1+1,\cdots,N_c-1 \tag{B.9}$$

If $\upsilon'_{m_1}$ and $\varpi'_{m_1}$ are sufficiently small, and $\upsilon'_{m_n}$ and $\varpi'_{m_n}$, for $m_n = 0,\cdots,m_1-1, m_1+1,\cdots,N_c-1$, are sufficient large, the inequality in (B.9) will hold. If the upper and lower bounds are sufficiently separated, some $\upsilon_{m_n}$ and $\varpi_{m_n}$, for $m_n = 0,1,\cdots,N_c-1$ will satisfy the condition in (B.9), since $\upsilon_{m_n}$ and $\varpi_{m_n}$, for $m_n = 0,1,\cdots,N_c-1$ are limited by the lower and upper bounds of the uncertainty classes.

**REFERENCE**

[1] A. Hassanien, M. G. Amin, Y. D. Zhang, and F. Ahmad, "Signaling strategies for dual-function radar communications: an overview," *IEEE Aerospace and Electronic Systems Magazine*, vol. 31, no. 10, pp. 36-45, Oct. 2016.

[2] A. Hassanien, M. G. Amin, Y. D. Zhang, and F. Ahmad, "Dual-function radar-communications: information embedding using sidelobe control and waveform diversity," *IEEE Transactions on Signal Processing*, vol. 64, no. 8, pp. 2168-2181, Apr. 2016.

[3] B. Paul, A. R. Chiriyath, and D. W. Bliss, "Survey of RF communications and sensing convergence research," *IEEE Access*, vol. 5, pp. 252-270, Feb. 2017.

[4] G. O. Young, G. Tavik, C. Hilterbrick, J. Evins, J. Alter, J. Crnkovich Jr., J. de Graaf, W. Habicht II, G. Hrin, S. Lessin, D. Wu, and S. Hagewood, "The advanced multifunction RF concept," *IEEE Transactions on Microwave Theory and Techniques*, vol. 53, no. 3, pp. 1009–1020, Mar. 2005.

[5] L. Han and K. Wu, "Multifunctional transceiver for future intelligent transportation systems," *IEEE Transactions on Microwave Theory and Techniques*, vol. 59, no.7, pp. 1879-1892, Jul. 2011.

[6] J. Moghaddasi and K. Wu, "Multifunctional transceiver for future radar sensing and radio communicating data-fusion platform", *IEEE Access*, vol. 4, pp. 818-838, Feb. 2016.

[7] C. Sturm and W. Wiesbeck, "Waveform design and signal processing aspects for fusion of wireless communications and radar sensing," *Proceedings of the IEEE*, vol. 99, no. 7, pp. 1236-1259, Jul. 2011.

[8] R. A. Romero and K. D. Shepherd, "Friendly spectrally shaped radar waveform with legacy communication systems for shared access and spectrum management," *IEEE Access*, vol. 3, pp. 1541-1554, Aug. 2015.

[9] H. Takase and M. Shinriki, "A dual-use radar and communication system with complete complementary codes," in *2014 15th International Radar Symposium (IRS)*, Gdansk, Jun. 16-18, 2014, pp. 16-18.

[10] R. M. Mealey, "A method for calculating error probabilities in a radar communication system," *IEEE Transaction on Space Electronics and Telemetry*, vol. 9, no. 2, pp. 37-42, Jun. 1963.

[11] S. D. Blunt, P. Yatham, and J. Stiles, "Intrapulse radar-embedded communications," *IEEE Transactions on Aerospace. Electronic Systems*, vol. 46, no. 3, pp. 1185-1200, Jul. 2010.

[12] Y. L. Sit and T. Zwick, "MIMO OFDM radar with communication and interference cancellation features," in *2014 IEEE Radar Conference*, Cincinnati, OH, May 19-23, 2014, pp. 19-23.

[13] Y. Liu, G. Liao, Z. Yang, and J. Xu, "Joint range and angle estimation for an integrated system combining MIMO radar with OFDM communication", *Multidimensional Systems and Signal Processing*, pp. 1-27, 2018.

[14] Y. Liu, G. Liao, and Z. Yang, "Range and angle estimation for MIMO-OFDM integrated radar and communication systems," in *2016 CIE International Conference on Radar*, Guangzhou, China, Oct. 10-13, 2016.

[15] Y. Liu, G. Liao, Z. Yang, J. Xu, "Design of integrated radar and communication system based on MIMO-OFDM waveform," *Journal of Systems Engineering and Electronics*, vol. 28, no. 4, pp. 669-680, Aug. 2017.

[16] W. Zheng, "Effects of power amplifier distortion and channel estimation errors on the performance of DVB-H system with multiple-antenna receiver," *IEEE Transactions on Consumer Electronics*, vol. 55, no. 4, pp. 1810-1818, Nov. 2009.

[17] H. Lin, R. C. Chang, K. Lin, and C. Hsu, "Implementation of synchronization for 2×2 MIMO WLAN system," *IEEE Transactions on Consumer Electronics*, vol. 52, no. 3, pp. 776-773, Aug. 2006.

[18] P. Liu, S. Wu, and Y. Bar-Ness, "A phase noise mitigation scheme for MIMO WLANs with spatially correlated and imperfectly estimated channels," *IEEE Communications Letters*, vol. 10, no. 3, pp. 141-143, Mar. 2006.

[19] M. Jankiraman, B. J. Wessels, and P. van Genderen, "Design of a multifrequency FMCW radar," in *28th European Microwave Conference*, Amsterdam, Netherlands, Oct. 1998, pp. 584-589.

[20] N. Levanon, "Multifrequency complementary phase-coded radar signal," *IEE Proceedings Radar, Sonar and Navigation*, vol. 147, no. 6, pp. 276–284, Dec. 2000.

[21] T. X. Zhang, X. G. Xia, and L. J. Kong, "IRCI free range reconstruction for SAR imaging with arbitrary length OFDM pulse," *IEEE Transactions on Signal Processing*, vol. 62, no. 18, pp. 4748-4759, Sep. 2014.

[22] T. X. Zhang and X. G. Xia, "OFDM synthetic aperture radar imaging with sufficient cyclic prefix," *IEEE Transactions on Geoscience and Remote Sensing*, vol. 53, no.1, pp. 394-404, Jan. 2015.

[23] Y. Cao, and X. Xia, "IRCI free MIMO-OFDM SAR using circularly shifted Zadoff-Chu sequences", *IEEE Geoscience and Remote Sensing Letters*, vol. 12, no.5, pp. 1126–1130, May 2015.

[24] S. Sen and A. Nehorai, "Adaptive OFDM radar for target detection in multipath scenarios," *IEEE Transactions on Signal Processing*, vol.59, no.1, pp. 78-90, Jan. 2011.

[25] S. Sen and A. Nehorai, "OFDM MIMO radar with mutual-information waveform design for low-grazing angle tracking," *IEEE Transactions on Signal Processing*, vol. 58, no. 6, pp. 3152–3162, Jun. 2010.

[26] S. Sen, "PAPR-constrained Pareto-optimal waveform design for OFDM-STAP radar," *IEEE Transactions on Geoscience and Remote Sensing*, vol. 52, no. 6, pp. 3658-3669, Jun. 2014.

[27] D Jiang, and L. Delgrossi, "IEEE 802.11p: Towards an international standard for wireless access in vehicular environments," In *IEEE Vehicular Technology Conference*, Singapore, May 11-14, 2008, pp. 2036-2040.

[28] IEEE 802.11 Working Group, "Part11: Wireless LAN medium access control (MAC) and physical layer (PHY) specifications," ANSI/IEEE Std. 802.11, 1999.

[29] "IEEE Std. 802.11-2007, Part 11: Wireless LAN Medium Access Control (MAC) and Physical Layer (PHY) specifications," IEEE Std. 802.11, 2007.

[30] A. D. Harper, J. T. Reed, J. L. Odom, and A. D. Lanterman, "Performance of a joint radar-communication system in doubly-selective channels," in *2015 49th Asilomar Conference on SSC*, Nov 8-11, 2015, pp. 1369-1373.

[31] A. R. Chiriyath, B. Paul, G. M. Jacyna, and D. W. Bliss, "Inner bounds on performance of radar and communications co-existence," *IEEE Transactions on Signal Processing*, vol. 53, no. 2, pp. 464-474, Feb. 2015.

[32] A. R. Chiriyath, B. Paul, and D. W. Bliss, "Radar-communications convergence: coexistence, cooperation, and co-design," *IEEE Transactions on Cognitive Communications and Networking*, vol. 3, no. 1, pp. 1-12, Mar. 2017.

[33] Y. Liu, G. Liao, Z. Yang, and J. Xu, "Multiobjective optimal waveform design for OFDM integrated radar and communication systems," *Signal Processing*, 141, pp. 331–342, 2017.

[34] Y. Liu, G. Liao, J. Xu, Z. Yang, L. Huang, and Y. Zhang, "Transmit power adaptation for orthogonal frequency division multiplexing integrated radar and communication systems," *Journal of Applied Remote Sensing*, vol. 11, no. 3, pp. 1-17 Sep. 2017.

[35] M. R. Bell, "Information theory and radar waveform design," *IEEE Transactions on Information Theory*, vol. 39, no. 5, pp. 1578-1597, Sep. 1993.

[36] Y. Yang and R. S. Blum, "MIMO radar waveform design based on mutual information and minimum mean-square error estimation," *IEEE Transactions on Aerospace and Electronic Systems*, vol. 43, no. 1, pp. 330-343, Jan. 2007.

[37] A. Leshem, O. Naparstek, and A. Nehorai, "Information theoretic adaptive radar waveform design for multiple extended targets," *IEEE Journal of Selected Topics in Signal Processing*, vol. 1, no. 1, pp. 42-55, Jun. 2007.

[38] U. Madhow, "Fundamentals of digital communication," New York: Cambridge University Press, 2008.

[39] B. Luo, Q. Cui, H. Wang, and X. Tao, "Optimal joint water-filling for coordinated transmission over frequency-selective fading channels," *IEEE Communications Letters*, vol. 15, no. 2, pp. 190-192, Feb. 2011.

[40] D. P. Palomar and J. R. Fonollosa, "Practical algorithms for a family of waterfilling solutions," *IEEE Transactions on Signal Processing*, vol. 53, no. 2, pp. 686-695, Feb. 2005.

[41] R. Xu, L. Peng, W. Zhao, and Z. Mi, "Radar mutual information and communication channel capacity of integrated radar-communication system using MIMO", *ICT Express*, vol. 1, no.3, 2015, pp.102-105.

[42] Y. Liu, G. Liao, J. Xu, Z. Yang, and Y. Zhang, "Adaptive OFDM integrated radar and communications waveform design based on information theory," IEEE Communications Letters, vol. 21, no. 10, pp. 2174-2177, Oct. 2017.

[43] W. Zhu, and J. Tang, "Robust design of transmit waveform and receive filter for colocated MIMO radar," *IEEE Signal Processing Letters*, vol. 22, no. 11, pp. 2112-2116, Nov. 2015.

[44] E. Grossi, M. Lops, and L.Venturino, "Robust waveform design for MIMO radars," *IEEE Transactions on Signal Processing*, vol. 59, no. 7, pp. 3262- 3271, Jul. 2011.

[45] Y. Yang, and R. S. Blum, "Minimax robust MIMO radar waveform design," *IEEE Journal Of Selected Topics in Signal Processing*, vol. 1, no. 1, pp. 1-9, Jun. 2007.

[46] S. A. Kassam, and H. V. Poor, "Robust signal processing for communication systems," *IEEE Communications Magazine*, vol. 73, no. 3, pp. 20-28, Jan. 1983.

[47] S. A. Kassam, and H. V. Poor, "Robust techniques for signal processing : a survey," *Proceedings of the IEEE*, vol. 73, no. 3, pp. 433-481, Mar. 1985.

[48] C. Williams, "Robust chaotic communications exploiting waveform diversity. Part 2: Complexity reduction and equalisation," *IET Communications*, vol. 2, no. 10, pp. 1223–1229, 2008.

[49] E. Nelson, "Radically elementary probability theory," Princeton University Press, 1987.

[50] S. Boyd and L. Vandenberghe, "Convex optimization," New York: Cambridge University Press, 2004.

[51] S. H. Han, J. H. Lee, "An overview of peak-to-average power ratio reduction techniques for multicarrier transmission," *IEEE Wireless Communications*, vol. 12, no. 2, pp. 56–65, Apr. 2005.

[52] J. G. Proakis, M. Salehi, "Digital Communications, Fifth Edition", McGraw-Hill, New York, 2008.